\documentclass[twocolumn,floats,showpacs,amsmath,amssymb,pre]{revtex4}
\usepackage{graphicx}
\begin{document}

\title{Diffusion of hydrogen in graphite: A molecular dynamics simulation}
\author{Carlos P. Herrero}
\author{Rafael Ram\'{\i}rez}
\affiliation{Instituto de Ciencia de Materiales de Madrid,
         Consejo Superior de Investigaciones Cient\'{\i}ficas (CSIC),
         Campus de Cantoblanco, 28049 Madrid, Spain }
\date{\today}

\begin{abstract}
Diffusion of atomic and molecular hydrogen in the interstitial space
between graphite sheets has been studied by molecular dynamics
simulations.
Interatomic interactions were modeled by a tight-binding potential
fitted to density-functional calculations.
Atomic hydrogen is found to be bounded to C atoms, and its diffusion 
consists in jumping from a C atom to a neighboring one, with an 
activation energy of about 0.4 eV.
Molecular hydrogen is less attached to the host sheets and diffuses 
faster than isolated H.
At temperatures lower than 500 K, H$_2$ diffuses with an activation
energy of 89 meV, whereas at higher $T$ its diffusion is enhanced by
longer jumps of the molecule as well as by correlations between 
successive hops, yielding an effective activation energy of 190 meV.
\end{abstract}


\maketitle

\section{Introduction}

The past few years have seen extraordinary progress in the knowledge 
of carbon-based materials, and in particular on those formed by C 
atoms displaying $sp^2$ hybridization. This is the case of
materials discovered in last decades, such as carbon nanotubes
and fullerenes, or in last years, such as graphene \cite{ge07,ka07},
apart from the more traditional graphite.

Carbon-based systems, in general, are considered as candidates for
hydrogen storage \cite{di01,ko05}. Moreover, chemisorption on 
two-dimensional systems, such as graphene or graphite surfaces, 
can be important for catalytic processes \cite{sl03}.
The interest on hydrogen as an impurity in solids and on surfaces is not 
new, and dates back to several decades. This is in principle
one of the simplest impurities, but a deep understanding 
of its physical properties is complex due to its low mass, and 
requires the combination of advanced experimental and theoretical 
methods \cite{pe92,es95}.
In addition to its basic interest as an impurity, a relevant
characteristic of hydrogen in solids and surfaces is its ability to 
form complexes and passivate defects, which has been extensively studied 
in the last thirty years \cite{pe92,es95,ze99}. 

 Experimental investigations on atomic, isolated hydrogen in graphite have 
been so far scarce, due to the difficulty in detecting this impurity. 
Moreover, this problem is complicated by the presence of a large
amount of hydrogen trapped at the boundaries of graphite 
crystallites \cite{at02,wa04,wa07}. 
The stable hydrogen configurations in the bulk of graphite have been
investigated in various theoretical works \cite{fe02,di03,sh03},
with particular emphasis on atomic and molecular forms of this
impurity.
Also, the diffusion, trapping, and recombination of hydrogen on a 
graphite surface have been studied by theoretical 
techniques \cite{fe03,fe04,sh05b,mo08}. 
In connection with this, atomic hydrogen on graphene has been
investigated by several authors using {\em ab-initio} 
methods \cite{sl03,ya07,ca09,bo08,du04,an08}.
It is generally accepted that chemisorption of a single hydrogen atom 
leads to the appearance of a defect-induced magnetic moment on the graphene 
sheet, along with a large structural distortion \cite{ya07,ca09,bo08}.
Recently, Ohldag {\em et al.} \cite{oh09} have discussed the role
of hydrogen in room-temperature ferromagnetism at graphite surfaces
from an x-ray dichroism analysis.

For the storage of hydrogen in graphite one should consider, apart
from atomic hydrogen, the presence of H$_2$ molecules between the
graphite sheets, which are expected to be physisorbed in the
interlayer space \cite{di03,fe02,at02}.
To study the diffusion of H$_2$ in the bulk of graphite, one has to
take into account that most of the experimental measurements detect
in fact the molecular diffusion through crystallite
boundaries \cite{at02}, so that it is hard to obtain direct insight
into the molecular diffusion in the interlayer space.

In this paper, molecular dynamics (MD) simulations are used 
to investigate the diffusion of atomic and molecular hydrogen in
graphite.
Recently, the diffusion of atomic hydrogen on an isolated graphene sheet
has been studied by a combination of path-integral molecular dynamics 
simulations and transition-state theory, with special emphasis 
upon the appearance of quantum effects \cite{he09}.
Here, we will deal with classical molecular dynamics
simulations, at temperatures high enough that such quantum effects
become unimportant.
This allows us to calculate directly diffusion coefficients, as the
time in the classical simulations corresponds to real time, contrary
to path-integral simulations, where the simulation time does not
strictly correspond to actual time, but appears as a convenient
computational way to derive time-independent thermodynamic properties.
In the simulations presented here, the interatomic interactions have 
been modeled by a tight-binding (TB) potential fitted to density-functional 
calculations.  
The thermal behavior of hydrogen between graphite layers, as well as
its diffusion in porous graphite have been addressed earlier by using
MD simulations \cite{sh03,wa04}.   
This computational technique has been also applied to study several
properties of hydrogen in various carbon-based
materials \cite{ro09,ha09,it09,ra09}.  In general, 
finite-temperature properties of hydrogen-related defects in solids
have been investigated by {\em ab-initio} and TB 
molecular dynamics simulations \cite{bu91,pa94,be00}. 

 The paper is organized as follows. In Sec.\,II, we describe the
computational method employed in our calculations. 
Our results are presented in Sec.\,III, dealing with the diffusion of
atomic and molecular hydrogen. 
In Sec.\,IV we summarize the main results.

\section{Computational Method}

 An important issue of the MD method is the adequate description 
of interatomic interactions, which should be as realistic as possible.
Since effective classical potentials present many limitations
to reproduce the many-body energy surface, specially in those situations
where interatomic bonds may be either broken or formed, one should 
resort to self-consistent quantum-mechanical methods. However,
density functional or Hartree-Fock based self-consistent potentials
require computer resources that would restrict enormously the size of
our simulation cell and/or the number of simulation steps. 
We found a compromise by employing an
efficient tight-binding  effective Hamiltonian, based on density
functional (DF) calculations \cite{po95}.
The ability of TB methods to reproduce different properties of
solids and molecules has been reviewed by 
Goringe {\em et al.} \cite{go97}
We checked the capability of this DF-TB potential to predict
frequencies of C--H vibrations. In particular, for a methane molecule 
it predicts in a harmonic approximation frequencies of 3100 and 
3242 cm$^{-1}$ for C--H modes with symmetry $A_1$ and $T_2$, 
respectively \cite{he06}, to be compared
with experimental values of 2917 and 3019 cm$^{-1}$ \cite{jo93}.
Considering the usual anharmonic shift (towards lower frequencies) 
associated to these modes, the agreement is satisfactory.
A detailed analysis of vibrational frequencies of hydrocarbon
molecules derived with the present DF-TB potential, including 
anharmonicities, can be found elsewhere \cite{lo03,bo01}.
We have employed earlier this TB Hamiltonian to describe 
hydrogen-carbon interactions in diamond \cite{he06,he07} and 
graphene \cite{he09}.  
The TB energy consists of two parts, the first one is the sum of
energies of occupied one-electron states, and the second one is
given by a pairwise repulsive interatomic potential \cite{po95}.
Since a correct description of the hydrogen molecule is essential
for our purposes, special care was taken with the H-H pair potential,
which has been taken as in our earlier study of molecular hydrogen in
the silicon bulk \cite{he09b}.  This pair potential reproduces 
the main features of known effective interatomic potentials for H$_2$, 
such as the Morse potential \cite{ra01}.

  Molecular dynamics simulations were carried out in the $NVE$ 
ensemble on a graphite supercell containing 64 C atoms and one
impurity (H or H$_2$), and periodic boundary conditions were assumed. 
The simulation cell includes two graphite sheets, each one being a 
$4\times4$ graphene supercell of size $4 a$ = 9.84 \AA.
We considered an $AB$ layer stacking, so that both sheets are disposed 
in such a way that the center of each hexagonal ring of one of them 
lies over a C atom of the adjacent sheet.  
To hold this kind of stacking along
a simulation run, thus avoiding diffusion of the graphite layers, the
center-of-gravity of each layer was not allowed to move on the 
$(x, y)$ plane. Note that in the following we will refer to the $z$
direction as the one perpendicular to the graphite layers.
The average distance between sheets is a half of the supercell parameter
along the $z$ axis, and was taken to be 3.35 \AA.
For the reciprocal-space sampling we have employed only the $\Gamma$
point (${\bf k} = 0$), as the main effect of using a larger ${\bf k}$
set is a nearly constant shift in the total energy, with little influence
in the calculation of energy differences.

Sampling of the configuration space has been performed
at temperatures between 300 and 2000 K. 
 For a given temperature, a typical run consisted of $3 \times 10^4$ MD 
steps for system equilibration, followed by $4 \times 10^6$ steps 
for the calculation of ensemble average properties. 
In some cases, specially at low temperatures, we carried out longer
simulations to reduce the statistical errors. In particular, for
H$_2$ at $T < 400$ K the simulations included $8 \times 10^6$ MD steps.
The equations of motion were integrated by using the standard Verlet 
algorithm \cite{ve67}.
The time step $\Delta t$ was taken in the 
range between 0.2 and 0.5 fs, which was found to be appropriate for the
atomic masses and temperatures studied here.
Some checks were carried out for smaller values of $\Delta t$,
yielding within error bars the same diffusion coefficients as those 
reported below.
The diffusion coefficient at different temperatures has been calculated
from the long-time behavior of the mean-square displacement of the 
mobile species (H or H$_2$) in the interlayer space.
Although motion along the $z$ coordinate (perpendicular to the
graphite sheets) is allowed, it does not contribute to the long-time
displacements, which are basically two-dimensional.
Thus, the diffusion coefficient is given by:
\begin{equation}
D = \frac{1}{4}   \lim_{t \to \infty} 
      \frac{(\Delta x(t))^2 + (\Delta y(t))^2}{t}  ,
\label{difcoef}
\end{equation}
where the mean-square displacement in the $x$ coordinate is
\begin{equation}
(\Delta x(t))^2 = \langle (x(t+t_0) - x(t_0) )^2 \rangle,
\end{equation}
and a similar expression holds for $(\Delta y(t))^2$.
Here $\langle ... \rangle$  means an average over different zero
times $t_0$ along a MD trajectory. 
For molecular hydrogen, the coordinates $x$ and $y$ correspond to the
center-of-mass of the molecule.

\section{Results}

\subsection{Atomic hydrogen}

We first discuss the lowest-energy configuration for atomic hydrogen 
in graphite, as derived from calculations at $T = 0$ K. 
The impurity binds to a C atom, which relaxes out of the sheet plane
by 0.34 \AA, with a bond distance between C and H of 1.17 \AA.
This result is in line with those reported in the literature,
with the breaking of a $\pi$ bond and the creation of
an additional $\sigma$ bond, changing the hybridization of the involved
C atom from $s p^2$ to $s p^3$ \cite{di03,fe03,sl03,bo08,ca09}.
This configuration with H attached to a C atom, which strongly relaxes
off the sheet plane, is similar to that found for hydrogen adsorbed
on an isolated graphene sheet \cite{sl03,bo08,he09}.

For its relation with the diffusion process, we have calculated
the energy barrier to break a C--H bond and attach the hydrogen
to a nearest C atom in the same sheet, and found an energy of 0.40 eV. 
Note that this value includes contributions of the appreciable 
relaxations of both carbon atoms involved in the whole process. 
It is interesting that we find the same energy barrier (within the
precision of our method) for hydrogen
to detach from a C atom in one sheet and jump to an opposite C atom 
in the adjacent sheet. This is not strange if one takes into account
that a hydrogen atom linked to a C atom in a graphite sheet is at
a distance of around 1.51 \AA\ from the sheet plane, a value
close to half the distance between graphite layers (1.67 \AA).
Since H moves towards the middle plane to brake a C--H bond, the
barriers for jumping to a C atom in the same or in the adjacent sheet
are similar. 
The energy value found here is close to those obtained in earlier DF
calculations: a value of 0.48 eV was reported in
Ref.~\cite{fe03}, and an estimated barrier of 0.4--0.5 eV
in Ref.~\cite{di03}.
We note that this barrier is clearly lower than the one found for
hydrogen jumps on an isolated graphene layer, for which the
interaction potential employed here gives an energy of 0.78 eV.

We now turn to the results of our MD simulations at finite temperatures.
From the calculations at $T$ = 0 K, one can expect that hydrogen will diffuse
between the graphite sheets in a step-like fashion, dissociating from
a C atom and attaching to a nearby one, in the same or in the adjacent
sheet. This will cause an almost two-dimensional diffusion in the
interlayer space. The likelihood of hydrogen jumping from an interlayer
region to an adjacent one, crossing a graphite sheet is very low, since
H has to climb an energy barrier of about 5 eV through an hexagonal
ring of C atoms.

\begin{figure}
\vspace{-2.0cm}
\hspace{-0.5cm}
\includegraphics[width= 9cm]{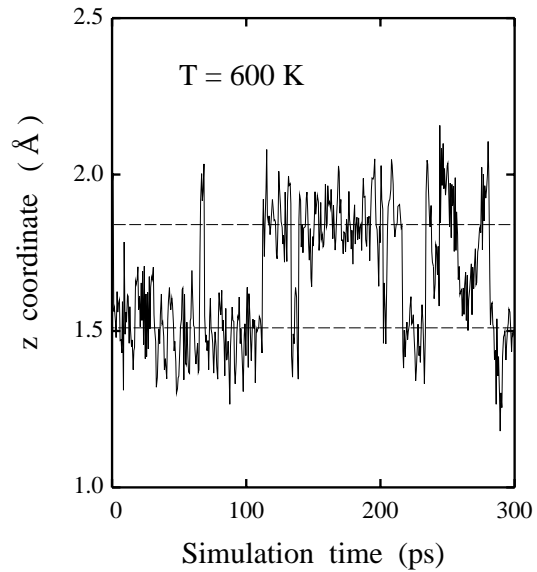}
\vspace{-2.5cm}
\caption{
Coordinate $z$ of hydrogen along a simulation run at 600 K.
The data shown include 10$^6$ MD steps, corresponding to a simulation
time of 300 ps.
The coordinate $z$ around 1.5 and 1.8 \AA\ corresponds to hydrogen
attached to C atoms either in the lower or in the upper graphite
sheet. $z$ is measured from the lower sheet.
}
\label{f1}
\end{figure}

In Fig.~1 we show the evolution of the $z$ coordinate of hydrogen 
(direction perpendicular to the graphite sheets) along a MD simulation
consisting of 10$^6$ steps (amounting to a time of 300 ps), at a
temperature of 600 K.
In spite of the large fluctuations in the coordinate $z$,
one observes that the hydrogen atom is mainly located on one of two 
planes at distance of about 1.5 or 1.8 \AA\ from the lower graphite layer.
These two distances correspond to H linked to C atoms in the lower and
upper sheets, respectively.
In fact, from the calculations at zero temperature, we found that the
stable sites for hydrogen (bound to C atoms) lie at 1.51 \AA\ from a
graphite sheet, and at 3.35 \AA\ - 1.51 \AA\ = 1.84 \AA\ from the 
adjacent one (see dashed lines in Fig.~1). 
In general, we observe along the MD simulations that the hydrogen 
jumps are rather short and direct, breaking a C--H bond and forming 
another one with a nearby carbon atom in the same or in an adjacent 
sheet.

\begin{figure}
\vspace{-1.8cm}
\hspace{-0.5cm}
\includegraphics[width= 9cm]{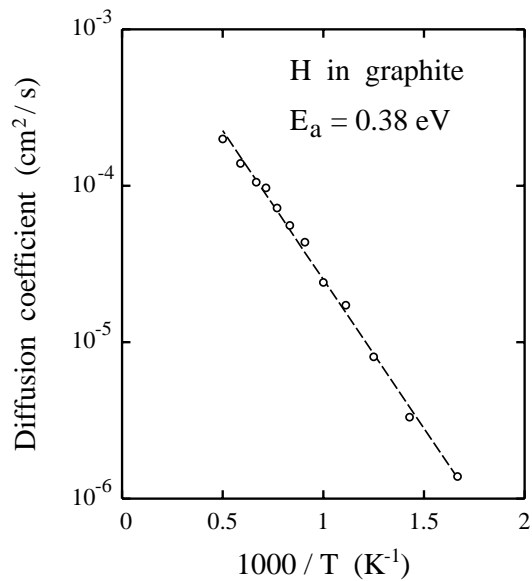}
\vspace{-2.5cm}
\caption{
Diffusion coefficient of atomic hydrogen in graphite, shown
in an Arrhenius plot vs the inverse temperature.
Error bars are on the order of the symbol size.
The dashed line is a least-square fit to the data points,
giving an effective activation energy $E_a$ = 0.38 eV.
}
\label{f2}
\end{figure}

From the mean-square displacement of hydrogen along the MD simulations, 
we have calculated the diffusion coefficient $D$ in the interlayer
space by using Eq.~(\ref{difcoef}).
The results are presented in Fig.~2 as a function of the inverse
temperature in an Arrhenius plot. They can be fitted well to an
expression $D = D_0 \exp(-E_a / k_B T)$, with an activation  energy
$E_a$ = 0.38 eV and a preexponential factor 
$D_0 = 2.0 \times 10^{-3}$ cm$^2$/s. 
This value corresponds to the order of magnitude expected for
H diffusion in graphite. In fact, $D_0$ can be expressed by the
simple expression: $D_0 = (\Delta r)^2 \nu_0 / 4$, where
$\Delta r$ is the distance between nearest adsorption sites (C atoms),
and $\nu_0$ is a typical ``attempt frequency'' (the factor of four in 
the denominator takes into account the fact that the diffusion is
two-dimensional, as in Eq.~(\ref{difcoef})).
Taking $\Delta r$ = 1.4 \AA\ and an attempt frequency corresponding to
a C--H stretching mode (about 3000 cm$^{-1}$), we find an estimation
for the preexponential factor $D_0 = 4.4 \times 10^{-3}$ cm$^2$/s.
This estimation is of the order of $D_0$ derived from the simulations, 
but its numerical value is somewhat larger. This is not strange if
one considers that hops of hydrogen from one layer to the adjacent one
can contribute $\Delta r$ = 1.4 \AA\ in the $(x, y)$ plane, but also
$\Delta r$ = 0 (jumping between opposite C atoms in the AB layer
stacking), without contributing to the overall diffusion.

At 1000 K we obtain from the MD simulations a relatively high 
diffusion coefficient
$D = 2.4 \times 10^{-5}$ cm$^2$/s. In practice, we can determine
rather accurately the value of $D$ down to temperatures in the order
of 600 K, where $D \sim 10^{-6}$ cm$^2$/s. At lower $T$, the 
hydrogen diffusion is too slow to allow for a precise determination of
$D$ from the mean-square displacements. In fact at room temperature 
($T$ = 300 K) hydrogen jumps are observed very rarely in the
simulations.
In any case, an extrapolation of the fit shown in Fig.~2 yields at
300 K a value $D =  9.2 \times 10^{-10}$ cm$^2$/s.
The activation energy derived from the slope of the dashed line
displayed in Fig.~2 ($E_a$ = 0.38 eV) is very close to the energy
barrier calculated for breaking a C--H bond and linking the hydrogen
to another C atom.
This confirms the diffusion mechanism of atomic hydrogen expected from
the calculations at $T$ = 0 K.

At $T$ lower than room temperature one expects the appearance 
of appreciable quantum effects in the hydrogen diffusion. 
Such quantum effects cause an effective renormalization of the 
diffusion barrier, which is reduced with respect to
the high-temperature value.  This can be studied by a combination of
transition-state theory with quantum path-integral simulations, as has
been done for hydrogen on a graphene sheet \cite{he09} and in 
bulk semiconductors \cite{he97,he07}, but is out of the scope of the
present work.

\subsection{Molecular hydrogen} 

\begin{figure}
\vspace{-2.0cm}
\hspace{-0.5cm}
\includegraphics[width= 9cm]{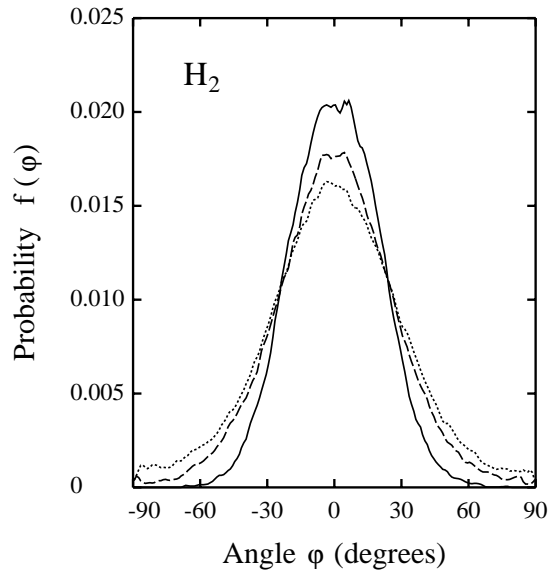}
\vspace{-2.5cm}
\caption{
Probability distribution $f(\varphi)$ for the angle $\varphi$
between the H--H direction and the graphite sheets, as derived
from MD simulations at three different temperatures: 300 K (solid
line), 800 K (dashed line), and 1300 K (dotted line).
}
\label{f3}
\end{figure}

Molecular hydrogen is expected to diffuse in graphite faster
than atomic hydrogen, since the former will be less bonded to the host
sheets, and therefore more mobile in the interlayer space. 
For an H$_2$ molecule we find a minimum-energy position at an
interstitial site between a carbon atom in a graphite sheet and an 
hexagonal ring in an adjacent sheet. At this position, the preferred
orientation of the molecule is parallel to the graphite planes, 
in agreement with earlier calculations based on density-functional
theory \cite{di03}.
At finite temperatures the molecule will explore other
positions and orientations with respect to the graphite layers.
We define the angular probability distribution $f(\varphi)$ of the
angle $\varphi$ between the H--H direction and the sheet plane
such that the probability $P(\varphi_1,\varphi_2)$ of observing
an angle $\varphi$ in the interval $(\varphi_1,\varphi_2)$ is
given by
\begin{equation}
P(\varphi_1,\varphi_2) = \int_{\varphi_1}^{\varphi_2} 
         f(\varphi) \cos \varphi \, d\varphi  
\end{equation}
(i.e., $\cos \varphi$ takes into account the degeneracy of angle
$\varphi$). 
In Fig.~3 we present the probability distribution $f(\varphi)$, as derived
from our molecular dynamics simulations at three different temperatures:
300 K (solid line), 800 K (dashed line), and 1300 K(dotted line).
The distribution has a maximum at $\varphi = 0$ (H--H parallel to the
layers), and reaches its minimum
for H--H perpendicular to the sheet plane ($\varphi = 90^o$).
As temperature increases, the probability distribution broadens
slightly, but it remains as a peak centered at $\varphi = 0$.

\begin{figure}
\vspace{-2.0cm}
\hspace{-0.5cm}
\includegraphics[width= 9cm]{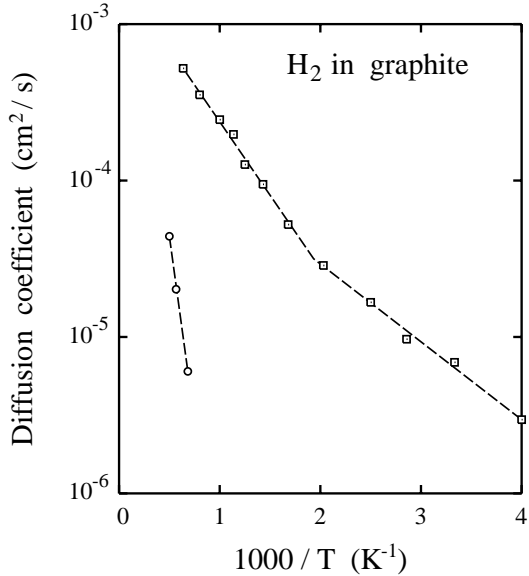}
\vspace{-2.5cm}
\caption{
Diffusion coefficient of molecular hydrogen in graphite as a function
of the inverse temperature, as derived from our MD simulations.
Squares represent data obtained from simulations where the C atoms
of graphite were allowed to move, whereas circles correspond to
simulations where the C atoms were kept fixed on their ideal equilibrium
sites in the graphite sheets.
Data shown as squares were fitted independently at high and low
temperatures, giving activation energies of 190 and 98 meV for
$T > 500$ K and $T < 500$ K, respectively.
Error bars of the simulation results are on the order of the symbol
size.
}
\label{f4}
\end{figure}

From the mean-square displacement of H$_2$ along the
molecular dynamics simulations we have calculated its diffusion
coefficient at several temperatures. The results are presented in Fig.~4
as a function of the inverse temperature. One observes first that the
diffusion coefficient of H$_2$ is larger than that of atomic
hydrogen in graphite, shown in Fig.~2. Thus, at 1000 K we find
$D = 2.5 \times 10^{-4}$ cm$^2$/s for H$_2$ to be compared with
$2.4 \times 10^{-5}$ cm$^2$/s for H.
This difference is much larger at 300 K, since we find
$D = 6.9 \times 10^{-6}$ cm$^2$/s for H$_2$ vs
the extrapolated value of $D =  9.2 \times 10^{-10}$ cm$^2$/s
for atomic hydrogen (about four orders of magnitude less).

We note that the results presented for the diffusion coefficient of H$_2$
were derived from MD simulations where the graphite sheets are flexible,
i.e. the C atoms are free to move along the simulations, and in
particular the sheets can relax and adapt to the presence of the
hydrogen molecule. 
However, due to the lack of direct bonding between molecular hydrogen
and graphite, one can ask it relaxation of the sheets does in fact
affect the molecular diffusion.
To obtain insight into this question we have carried out some MD
simulations in which the C atoms were kept fixed at their unrelaxed
(ideal) positions, and calculated the diffusion coefficient of H$_2$.
The results of these calculations are displayed in Fig.~4 as circles at
temperatures higher than 1000 K. One notices that, at a given $T$,
the value of $D$ derived in this way is clearly lower than that found 
when the C atoms are allowed to move and relax in the presence of 
H$_2$. In fact, at 1000 K the diffusion coefficient with fixed C atoms
is smaller than $10^{-6}$ cm$^2$/s, the lower limit for a reliable
determination of $D$ in our simulations.
All this indicates that motion and relaxation of the graphite sheets
directly affects the diffusion of molecular hydrogen in graphite.

A remarkable aspect of Fig.~4 is that the diffusion coefficient of
molecular hydrogen does not follow a single straight line in the
Arrhenius plot, i.e. it does not display a dependence of the form
$D = D_0 \exp(-E_a / k_B T)$ in the whole temperature range considered
here. It seems rather that one can distinguish two temperature 
regions with different effective activation energy $E_a$.
In fact, at $T > 500$ K those data can be fitted well with
$E_a$ = 0.19 eV, whereas at lower temperature $E_a$ seems to be about
a factor of two lower ($E_a$ = 0.098 eV).

\begin{figure}
\vspace{-2.0cm}
\hspace{-0.5cm}
\includegraphics[width= 9cm]{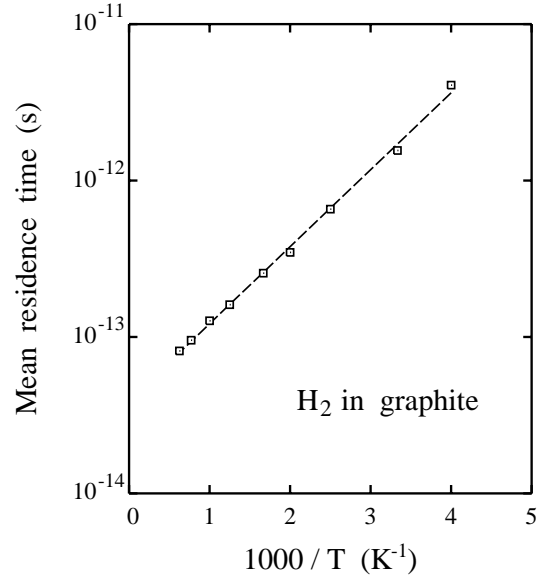}
\vspace{-2.5cm}
\caption{
Mean residence time of molecular hydrogen in an interlayer site of
graphite plotted vs the inverse temperature.
Error bars are in the order of the symbol size.
A least-square fit to the data points (dashed line) gives an activation
energy of 97 meV.
}
\label{f5}
\end{figure}

Given the change in slope of the results presented in the Arrhenius plot
of Fig.~4 (squares), we wonder whether at high and low temperatures the
diffusion mechanism is different, or maybe the molecular jump rate from site
to site in the interlayer space suffers some change apart from that
predicted by a single activation energy.
To clarify this question we have calculated the mean residence time of
H$_2$ on the interstitial sites along the MD simulations. To this end we
have counted the number of molecular jumps in the molecular dynamics
trajectories at different temperatures. The results for the mean
residence time $\tau$ are plotted in Fig.~5 vs the inverse temperature.
$\tau$ is found to follow a dependence $\tau = \tau_0 \exp(E_a/k_B T)$ 
in the whole temperature range under consideration, with an activation 
energy $E_a$ =  97 meV.
This activation energy coincides, within statistical error, with that
derived from the diffusion coefficient of molecular hydrogen at
temperatures lower than 500 K ($E_a$ = 98 meV).

The available sites for H$_2$ in the interlayer space of graphite form a
honeycomb lattice, in which each site has three nearest neighbors.
At low temperature ($T < 500$ K) the molecular diffusion proceeds via 
jumps between nearest-neighbor adsorption sites in a random walk of the
H$_2$ molecule in the interlayer space. At high temperatures, however,
we observe two mechanisms that contribute to enhance the molecular
diffusion. First, we detected jumps longer than the distance between
nearest interstitial adsorption sites. In particular, we found direct 
jumps to second (next-nearest) and third neighboring interlayer sites, 
at a distance of 1.7 and 2 times that between nearest neighbors.
 At 1000 K, the fraction of these longer jumps amounts to about
9\% of the total number of molecular jumps, versus around 1\% at
500 K, and a negligible quantity of 300 K.

At $T > 500$ K we also observed that the molecular jumps between nearest 
adsorption sites are correlated, in the sense that after a given jump,
the probability that the next one occurs in the forward direction
(the same direction as the previous one) is larger than in any other 
direction. This correlation contributes to enhance the diffusion of
H$_2$, and at high temperatures the molecular diffusion cannot be
considered as a random walk over the available adsorption sites.
These two mechanisms (longer and correlated jumps) cause an increase
in the diffusion coefficient, as compared with the behavior expected
from an extrapolation of the low-temperature results. Such an
increase translates into an enhancement of the slope in the Arrhenius
plot of the diffusion coefficient, and consequently in the observation 
of an {\it apparently} larger effective activation energy $E_a$.
This behavior illustrates the limitations of the interpretation of
diffusion coefficients in Arrhenius plots. In fact, any 
temperature-dependent effect in the preexponential factor cannot be 
directly distinguished in an Arrhenius plot from changes in the
activation energy. Our results of the mean residence time of H$_2$ in
graphite provide evidence that the energy barrier for molecular motion 
remains constant in the temperature range from 250 to 1600 K.

\section{Summary}

The main advantage of this kind of MD simulations of hydrogen in 
graphite is the possibility of calculating diffusion coefficients 
in a large range of temperatures, using an interatomic potential
fitted to {\em ab-initio} calculations.
Due to the large relaxation of the nearest C atoms, the migration 
of atomic hydrogen in graphite requires important motion of these atoms. 
Then, a hydrogen jump has to be viewed as a cooperative process
involving a coupled motion of the impurity and the nearest host atoms.  
This effect is not a requirement for the diffusion of 
H$_2$, but in practice relaxation of the C atoms in the 
nearest graphite layers helps to enhance appreciably the molecular
diffusion. This has been shown in Fig.~4 by comparing MD simulations 
with fixed or mobile carbon atoms. In fact, at 1000 K the diffusion 
coefficient is more than 100 times larger when the C atoms are
allowed to relax out of their ideal positions.

At 1000 K we obtained for molecular hydrogen a diffusion coefficient
one order of magnitude larger than for atomic hydrogen.
In fact, we found $D = 2.5 \times 10^{-4}$ cm$^2$/s for H$_2$ versus
$2.4 \times 10^{-5}$ cm$^2$/s for H.
This difference increases to about four orders of magnitude at
300 K: $D = 6.9 \times 10^{-6}$ cm$^2$/s for H$_2$, to be compared
with an extrapolated value of $D =  9.2 \times 10^{-10}$ cm$^2$/s

The diffusion coefficients derived here for atomic and molecular
hydrogen in graphite have to be considered as higher limits for the
actually measured values. In fact, point or extended defects in the
host layers will act as hydrogen traps, causing an appreciable 
reduction in the diffusion. More important, for a comparison with
experimental measurements, one has to take into account that an
important part of the hydrogen is concentrated on the crystallite
boundaries and in the micro-voids between graphite granules.

An interesting question is the dissociation of molecular hydrogen and
its recombination in the interlayer space of graphite. 
A study of the combination of these processes with atomic and 
molecular diffusion 
lies, however, out of the scope of the present paper and remains as a
challenge for future research.

\begin{acknowledgments}
This work was supported by Ministerio de Ciencia e Innovaci\'on (Spain) 
through Grants FIS2006-12117-C04-03 and FIS2009-12721-C04-04,
and by Comunidad Aut\'onoma de Madrid through Program
MODELICO-CM/S2009ESP-1691. 
\end{acknowledgments}

\end{document}